\documentclass[final]{aipproc}
\layoutstyle{6x9}

\usepackage{amsmath}
\usepackage{amssymb}
\usepackage{amscd}
\usepackage{fleqn}
\usepackage{graphics}
\usepackage{pifont}
\usepackage{latexsym}
\usepackage[small,rm]{caption}
\usepackage{graphicx}
\usepackage{boxedminipage}
\usepackage{pstricks}
\usepackage{pst-node}
\usepackage{multicol}
\usepackage{subfigure}
\usepackage{amsfonts}
\usepackage{amsthm}
\usepackage{fancyhdr}
\usepackage{dsfont}
\usepackage{bm,epic,eepic}
\usepackage[mathscr]{euscript}
\usepackage{booktabs} 
\usepackage{textcomp}
\usepackage{epsfig}
\usepackage{epstopdf}

\newcommand{\openone}{\leavevmode\hbox{\normalsize1\kern-3.8pt\large1}}
\newcommand{\BEQ}{\begin{eqnarray}}
\newcommand{\EEQ}{\end{eqnarray}}
\newcommand{\ENQ}{\end{eqnarray}}
\newcommand{\BE}{\begin{eqnarray}}
\newcommand{\EE}{\end{eqnarray}}
\newcommand{\bq}{\begin{quote}}
\newcommand{\eq}{\end{quote}}
\newcommand{\nn}{\nonumber}

\newcommand{\inhoud}[2]{\hbox to #1{\hss #2 \hss}}
\newcommand{\forget}[1]{}
\newcommand{\ket}[1]{| \, #1 \rangle}

\newcommand{\beq}{\begin{equation}}
\newcommand{\eeq}{\end{equation}}
\newcommand{\be}{\begin{equation}}
\newcommand{\ee}{\end{equation}}
\newcommand{\enq}{\end{equation}}

\newcommand{\1}{\mathds{1}}

\newcommand{\av}[1]{\langle #1 \rangle}

\renewcommand{\vec}[1]{\mathbf{#1}}

\renewcommand{\H}{\mathcal{H}}

\def\be{\begin{equation}}
\def\ee{\end{equation}}
\def\eea{\end{eqnarray}}
\def\bea{\begin{eqnarray}}

\begin{document}

\title{No-signaling, perfect bipartite dichotomic correlations and local randomness}


\author{M.P. Seevinck}{
  address={Institute for Mathematics, Astrophysics and Particle Physics, Faculty of Science \& Centre for the History of Philosophy and Science, Faculty of Philosophy, Radboud University Nijmegen, the Netherlands}
}

\begin{abstract}
The no-signaling constraint on bi-partite correlations is reviewed. It is shown that in order to obtain non-trivial Bell-type inequalities that discern no-signaling correlations from more general ones, one must go beyond considering expectation values of products of observables only. A new set of nontrivial no-signaling inequalities is derived which have a remarkably close resemblance to the CHSH inequality, yet are fundamentally different. A set of inequalities by \citet{roysingh} and \citet{avis}, which is claimed to be useful for discerning no-signaling correlations, is shown to be trivially satisfied by any correlation whatsoever. Finally, using the set of newly derived no-signaling inequalities a result with potential cryptographic consequences is proven: if different parties use identical devices, then, once they have perfect correlations at spacelike separation between dichotomic observables, they know that because of no-signaling the local marginals cannot but be completely random.
\end{abstract}

\maketitle


\section{I: Introduction}\noindent
Ever since Bell's seminal set of papers \cite{bell64,bell66} the study of non-local and quantum correlations has been of paramount importance for the understanding of the foundations of quantum physics. Recently, however, it has become clear that a different, more general kind of correlations need to be understood as well. These are the correlations that obey a no-signaling constraint, which is roughly the requirement by special relativity that signals cannot be communicated in a spacelike fashion. During the last ten years or so these no-signaling correlations have been extensively studied.

In this paper we first review in section II the idea of bi-partite correlations as joint probability distributions and what is known about the structure of the convex set of such probability distributions.  In more detail we will next consider the no-signaling constraint and its associated polytope of no-signaling correlations. We restrict ourselves to two parties only, but we note that generalisations are not straightforward, see \citet{thesis}. In section III it is shown that in order to obtain non-trivial Bell-type inequalities that discern no-signaling correlations from more general ones, one must go beyond considering expectation values of products of observables only (as is for example the case in the CHSH inequality for local correlations). A new set of nontrivial no-signaling inequalities is derived in section IV which have a remarkably close resemblance to the CHSH inequality, yet are fundamentally different. A set of inequalities by \citet{roysingh} and \citet{avis}, which is claimed by them to be useful for discerning no-signaling correlations, is shown to be trivially satisfied by any correlation whatsoever, including signaling ones. Finally, using the newly presented set of no-signaling inequalities a result with potential cryptographic consequences is proven: if different parties use identical devices then, once they have perfect correlations at spacelike separation between dichotomic observables, they know that because of no-signaling the local marginals cannot but be completely random. 
The importance of this result is discussed in the final section V.

\section{II: Correlations}\noindent
\label{allcorr}
\subsection{General correlations}\noindent
Consider $2$ parties, labeled by $1,2$, each holding a physical system that are to be measured using a finite set of different observables. Denote by $a$ the observable (random variable) that party $1$ chooses (also called the setting $a$)  and by $A$ the corresponding measurement outcome, similarly for $b$ and $B$ for party $2$. We assume there to be only a finite number of discrete outcomes. 
The outcomes can be correlated in an arbitrary way. 
A general way of describing this situation, independent of the underlying physical model, is by a set of
joint probability distributions for the outcomes, conditioned on the settings chosen by the $N$ parties, where the correlations  are captured in terms of these joint probability distributions. They are denoted by 
\begin{align}\label{generalcorr}
P(A,B|a,b).
\end{align}
These probability distributions are assumed to be positive 
\begin{align}\label{posgen}
P(A,B|a,b)\geq 0,~~ \forall A,B,~~ \forall a,b,
\end{align}
and obey the normalization conditions
\begin{align}\label{norma}
\sum_{A,B}P(A,B|a,b)=1,~~ \forall A,B,~~ \forall a,b.
\end{align}
We need not demand that the probabilities should not be greater than 1 because this follows from them being positive and from the normalization conditions.

The set of all these probability distributions has a nice structure. First, it is a convex set: convex combinations of correlations are still legitimate correlations. Second, there are only a finite number of extremal correlations. This means that every correlation can be  decomposed into a (not necessarily unique) convex combination of such extremal correlations. 

In general, a total of $D=m_{a} m_{A}m_{b}m_{B}$ different probabilities exist (here $m_{a}$ and $m_{A}$ are the number of different observables  and outcomes for party $1$ respectively, and similalry for party 2). When these conditional probability distributions (\ref{generalcorr}) are considered as points in a $D$-dimensional real space,  this set of points forms a convex polytope with a finite number of extreme points which are the vertices of the polytope. This polytope is  the convex hull of the extreme points. It  belongs to the subspace defined by (\ref{norma}) and it is bounded by a set of facets, linear inequalities that describe the halfplanes that bound it. Every convex polytope has a dual description, firstly in terms of its vertices, and secondly in terms of its facets, i.e., hyperplanes that bound the polytope uniquely. In general each facet can be described by linear combinations of joint probabilities which reach a maximum value at the facet, i.e., 
\begin{align}\label{facet}
\sum_{A,B;a,b} c_{A,B;a,b}P(A,B|a,b)\leq I,
\end{align}
with real coefficients $c_{A,B;a,b}$ and a real bound $I$ that is reached by some extreme points. 
 For each facet some extreme points of the polytope lie on this facet and thus saturate the inequality (\ref{facet}), while the other extreme points cannot violate it.   In general, when the extreme points are considered as vectors,  a hyperplane is a facet of a $d$-dimensional polytope iff $d$ affinely independent extreme points satisfy the equality that characterizes the hyperplane\footnote{\label{affine}In case the null vector belongs to the polytope, the condition of the existence of $d$ affinely independent vectors is equivalent to the existence of $(d-1)$ linearly independent vectors; otherwise it requires the existence of $d$ linearly independent vectors \cite{masanes02}.}. Consequently, for the case of general correlations    (\ref{generalcorr}) the set of extreme points that lie on a facet  must contain a total of $D$ affinely independent vectors.   For this case the facets are determined by equality in (\ref{posgen}).  The probability distributions  (\ref{generalcorr}) correspond to any normalized vector of positive numbers in this polytope.  For an excellent overview of the structure of these polytopes, see \cite{masanes02}, \cite{barrett05} and \cite{ziegler}.

The extreme points are the probability distributions that saturate a maximum of the positivity conditions (\ref{posgen}) while satisfying the normalization condition (\ref{norma}).  They are characterized by \citet{jones} to be the probability distributions such that for each set of settings $\{a,b\}$ there is a unique set of outcomes $ \{ A[a],B[b]\}$ for which $P(A,B|a,b)=1$,  with  $A[a]$  the deterministic determination of outcome $A$ given the setting $a$, etc. 
 There is thus a one-to-one correspondence between the extreme points and the sets of functions 
$\{A[a],B[b]\}$ from the settings to the outcomes. Any such set defines an extreme point.  The extreme points thus correspond to deterministic scenarios: each outcome is completely fixed by the totality of all settings and consequently there is no randomness left: $P(A,B|a,b)= \delta_{A,A[a,b]}\cdots  \delta_{B,B[a,b]}$. Finding all the facets of a polytope knowing its vertices is called the hull problem and this is in general a computationally hard task \cite{pitowsky}.  The facet descriptions (\ref{facet}) will be called Bell-type inequalities.

The marginal probabilities are obtained in the usual way from the joint probabilities by summing over the outcomes of the other parties. It is important to realize that for general correlations these marginals may depend on the settings chosen by the other parties. For example, in the case of two parties that 
each choose two possible settings $a,a'$ and $b,b'$ respectively, the marginals for party 1 are given by 
\begin{subequations}
\label{M1}
\begin{align}
P(A|a)^{b}&:= \sum_{B}P(A,B|a,b),\label{m1}\\
P(A|a)^{b'}&:=\sum_{B}P(A,B|b,b'),\label{m2}
\end{align}
\end{subequations}
and analogously for setting $a'$ and for the marginals of party 2. The marginal $P(A|a)^{b}$ may thus in general be different from  $P(A|a)^{b'}$.
 
We will now put further restrictions besides normalization and posivity on the probability distributions  (\ref{generalcorr})  that are motivated by physical considerations. We will here not be concerned with arguing for the plausibility of these physical considerations, nor what 
violations of these physically motivated restrictions amounts to, but merely give the definitions that will be used in future sections.

\subsection{No-signaling correlations}\label{nosigsec}\noindent
A no-signaling correlation for two parties is a correlation such that party $1$ cannot signal to party $2$ by the choice of what observable is measured by party $1$ and vice versa. This means that the marginal probabilities $P(A|a)^{b}$ (see \eqref{m1}) and $P(B|b)^{a}$ are independent of $b$ and $a$ respectively: 
\begin{subequations}
\label{22nosig}
\begin{align}
P(A|a)^{b}=P(A|a)^{b'}&:=P(A|a)
\forget{&\Longleftrightarrow \sum_{B}P(A,B|a,b)=\sum_{B}P(A,B|a,b'):=P(A|a)},~~ \forall A,a,b,b',\\
P(B|b)^{a}=P(B|b)^{a'}&:=P(B|b)
\forget{\\
&\Longleftrightarrow\sum_{A}P(A,B|a,b)=\sum_{A}P(A,B|a',b):=P(B|b)},~~ \forall B,a,a',b.
\end{align}
\end{subequations}
In a no-signaling context the marginals can thus be defined as $P(A|a)$, etc., i.e., without any dependence on far-away settings\footnote{Loubenets \cite{loubenets} claims that this latter claim is not true in general, and that one should distinguish the requirement $P(A|a)^{b}=P(A|a)^{b'}$ from $P(A|a)^{b}=P(A|a)$, where the first is called `no-signaling' and the second `EPR locality'. However, if one quantifies over all $b$ and $b'$ --as one should-- these conditions in fact become identical.}. 
These linear equations \eqref{22nosig} characterize an affine set \cite{masanes06}. The intersection of this set with the polytope of distributions 
(\ref{generalcorr}) gives another convex polytope: the no-signaling polytope. 
The vertices of this polytope can be split into two types: vertices that correspond to deterministic scenarios,  where all probabilities are either $0$ or $1$, and those that correspond to non-deterministic scenarios.  All no-signaling deterministic correlations are in fact local \cite{masanes06}, i.e., they can be written in terms of the local correlations that will be defined below. But all non-deterministic vertices correspond to non-local scenarios.

The facets of the no-signaling polytope follow from the defining conditions for no-signaling correlations. These are thus the trivial facets that follow from the positivity conditions as well as the non-trivial ones that follow from the no-signaling requirements  \eqref{22nosig}. The importance of the non-trivial facets of the no-signaling polytope is that if a point, representing some experimental data, lies within the polytope, then a model that uses no-signaling correlations exists that reproduces the same data. On the contrary, if the point lies outside the polytope and thus violates some Bell-type inequality describing a facet of the no-signaling polytope, then the data cannot be reproduced by a no-signaling model only, i.e., including signaling strategy is necessary. 
\\\\
%

   \subsection{Local correlations} \label{sectionlocal}\noindent
Local correlations are those that can be obtained if the parties are non-communica-ting and share classical information, i.e.,  they only have local operations and local hidden variables (also called shared randomness) as a resource. We take this to mean that these correlations can be written as
\begin{align}\label{localdistr}
P(A,B|a,b)=\int_\Lambda d\lambda p(\lambda) P(A|a,\lambda)  P(B|b,\lambda),
\end{align}
where $\lambda\in\Lambda$ is the value of the shared local hidden variable, $\Lambda$ the space of all hidden variables and $p(\lambda)$ is the probability that a particular value of $\lambda$ occurs. Note that $p(\lambda)$ is independent of the outcomes $A,B$ and settings $a,b$. This is a `freedom' assumption, i.e., the settings are assumed to be free variables. Furthermore, $P(A|a,\lambda)$ is the probability that outcome $A$ is obtained by party $1$ given that the observable measured was $a$ and the shared hidden variable was $\lambda$, and similarly for the other terms.  Since these probabilities are conditioned on the hidden variable $\lambda$ we will call them subsurface probabilities, in contradistinction to the probabilities $P(A|a)$, etc., that only conditionalize on the settings, which we call surface probabilities.

Condition (\ref{localdistr})  is supposed\footnote{Opinions differ on how to motivate (\ref{localdistr}), see e.g. \citet{seevuff10}. The technical results of this paper do not depend on such a motivation and whether it is physically plausible and/or sufficient.} to capture the idea of locality in a hidden-variable framework and it is called Factorisability, and models that give only  local correlations are called local hidden-variable (LHV) models.  Correlations that cannot be written as (\ref{localdistr}) are called non-local.  Local correlations are no-signaling thus the marginal probabilities derived from local correlations are defined in the same way as was done for no-signaling correlations, cf. \eqref{22nosig}.   

Let us review what is known about the set of local correlations. First, it is also a polytope with vertices (extremal points) corresponding to local deterministic distributions \cite{werwolf}, i.e., $P(A,B|a,b)= \delta_{A,A[a]}\delta_{B,B[b]}$ where  the function $A[a]$ gives the deterministic determination of outcome $A$ given the setting $a$, etc. Thus for each set of settings $\{a,b\}$ there is a unique set of outcomes $ \{ A[a],B[b]\}$ for which $P(A,B|a,b)=1$.  
  All these vertices are also vertices of the no-signaling polytope \cite{barrett05}. The  local polytope is known  to be constrained by two kinds of facets \cite{werwolf}. The first are trivial facets and derive from the positivity conditions (\ref{posgen}). Note that these are also trivial  facets of the no-signaling polytope. The second kind of facets are non-trivial and can be violated by non-local correlations. These are not facets of the no-signaling polytope. All facets are mathematically described by Bell-type inequalities \eqref{facet} that will be further introduced below.   Determining whether a point lies within the local polytope, i.e., whether it does not violate a local Bell-type inequality, is in general very hard as \citet{pitowsky} has shown this to be related to some known hard problems in computational complexity (i.e., it is an NP-complete problem). Furthermore, determining whether a given inequality is a facet of the local polytope is of similar difficulty (i.e., this problem is co-NP complete \cite{pitowsky91}).

\subsection{Quantum correlations}\label{qmcorrsection}\noindent
Lastly, we consider another class of correlations: those that are obtained by general measurements on quantum states (i.e.,  those that can be generated if the parties share quantum states). These can be written as 
\begin{align}\label{quantumcorre}
P(A,B|a,b)=\textrm{Tr}[M_{A}^{a} \otimes\cdots\otimes M_{B}^{b} \rho].
\end{align}
Here $\rho$ is a quantum state (i.e., a unit trace semi-definite positive operator) on a Hilbert space $\H=\H_1\otimes\H_2$, where $\H_1$ is the quantum state space of the system held by party $1$, and similarly for party 2. The sets $\{M_{A}^{a},M_{B}^{b} \}$ define what is called a positive operator valued measure (POVM), i.e., a set of positive operators $\{M_{A}^{a}\}$ satisfying 
$\sum_{A}M_{A}^{a}=\1,\forall a$.  
 Note that \eqref{quantumcorre} is linear in both $M_{A}^{a}$, $M_{B}^{b} $ and $\rho$, which is  a crucial feature of quantum mechanics.

Quantum correlations are no-signaling and therefore the marginal probabilities derived from such correlations are defined in the same way as was done for no-signaling correlations (cf. \eqref{22nosig}).  For example, the marginal probability for party $1$ is given by $P(A|a)=\textrm{Tr}[M_{A}^{a}\rho^1]$, where $\rho^1$ is the reduced state for party 1.

The set of quantum correlations has been investigated by, e.g., \citet{pitowsky} and \citet{wernerwolf2},
and is shown to be convex\forget{\footnote{This set is known to be convex for the case of all possible measurements on all possible quantum states \cite{pitowsky,wernerwolf2}. However, convexity is not proven if one restricts oneself to the measurements on a given state and to projector valued measurements (PVM) on a Hilbert space with given dimension.}}. It is not a polytope because the number of extremal points is not finite and consequently it has an infinite number of bounding halfplanes. Therefore we will refer to this set as the quantum body, in contradistinction to the sets of the other types of correlations which are referred to as polytopes.

\section{III: On comparing and discriminating the different kinds of correlations}\label{comparecorr}\noindent
The polytope of general correlations strictly contains the no-signaling polytope, which in turn contains the quantum body, which in turn contains the local polytope.  These results are obtained by comparing the facets of the relevant no-signaling and local polytopes with the halfplanes that bound the quantum bodies.  These facets  (i.e., bounding hyperplanes in the case of quantum correlations) are of course implicitly determined by the defining restrictions on the different types of correlations, but to find explicit experimentally accessible expressions for them is a hard job. A fruitful way of doing so is using so-called Bell-type inequalities. This will be discussed next.

\subsection{Bell-type inequalities}\noindent
\label{techbellineq}
We will restrict ourselves to Bell-type inequalities for the case where each party chooses between two alternative observables and where each observable is dichotomic, i.e., the observable has two possible outcomes which we denote by $\pm1$. 

Bell-type inequalities denote a specific bound on a linear sum of joint probabilities as in \eqref{facet}. The bound is characteristic of the type of correlation under study. However, frequently they are formulated not in terms of probabilities but in terms of product expectation values\footnote{These are also known as `joint expectation values' or `correlation functions', but we will not use this terminology.}, e.g., expectation values of products of observables $a$ and $b$, which we will denote by $ \av{ab}$. These are defined in the usual way as the weighted sum of the products of the outcomes:
 \begin{align}\label{expectation}
\av{ab}:=\sum_{A,B} AB~ P(A,B|a,b).
\end{align}
 Since we are restricting ourselves to dichotomic observables with outcomes $\pm1$ all expectation values are bounded by: $-1\leq\av{ab}\leq 1$, for all $a,b$.

The probabilities $P(A,B|a,b)$  in (\ref{expectation}) are determined using the different kinds of correlations we have previously defined.  If they are of the no-signaling, local or quantum form  
we denote the product expectation values they give rise to by $\av{ab}_{\textrm{ns}}$, $\av{ab}_{\textrm{lhv}}$ or $\av{ab}_{\textrm{qm}}$ respectively.

Very often the different possible correlations are investigated using Bell-type inequalities in terms of product expectation values instead of directly in terms of the joint probabilities.  
The main reason for this is that using the product expectation values simplifies the investigation considerably.  For example, consider the case of two parties that each measure two dichotomous observables each. We denoted them as $a,a'$ and $b,b'$ respectively, with outcomes $A,A'$ and $B,B'$.  Instead of dealing with the $16$-dimensional space of vectors  with components 
$P(A,B|a,b), P(A',B|a,b), \ldots, P(A',B'|a',b')$ we only have to deal with  the $4$-dimensional space of vectors that have as components the quantities $\av{ab}$,$\av{ab'}$,$\av{a'b}$,$\av{a'b'}$. To transform a vector from the $16$-dimensional space  to its corresponding $4$-dimensional space, one needs to perform a projection as given in (\ref{expectation}). It is known that the projection of a convex polytope is always a convex polytope \cite{masanes02}. Therefore, the convex polytopes we have considered previously for general, no-signaling, partially-local and local correlations in the higher dimensional joint probability space correspond to convex polytopes in the lower dimensional space of product expectation values. The set of vectors with components $\av{ab}$,$\av{ab'}$,$\av{a'b}$,$\av{a'b'}$  that are attainable by general, no-signaling, partially-local and local correlations are thus also characterized by a finite set of extreme points and corresponding facets. 

Dealing with the expectation values $\av{ab}$ is much simpler than dealing with the joint probabilities 
$P(AB|ab)$, although in general, the projection (\ref{expectation}) is not structure preserving and some information about the correltions might get lost. For example some non-local correlations could be projected into locally achievable expectation values of products of observables. But for the case of two parties that  each choose two dichotomous observables, as in the set-up of the CHSH inequality, this does not happen. Indeed,  in the next subsection we will see that the CHSH inequalities describe all non-trivial facets of the local polytope.
  The $4$-dimensional vectors with components $\av{ab}$,$\av{ab'}$,$\av{a'b}$,$\av{a'b'}$  and the $16$-dimensional vectors  with components $P(A,B|a,b), P(A',B|a,b), \ldots, P(A',B'|a',b')$ thus contain the same information concerning the existence of a LHV model accounting for them.

\subsection{Dichotomic example: the CHSH inequality}\label{chsintrotech}\noindent
The best-known Bell-type inequality is the CHSH inequality for local correlations \cite{chsh} that assumes a situation of two parties and two dichotomous observables per party (with possible outcomes $\pm1$). 
Consider the CHSH polynomial:
\begin{align} \label{CHSHpolynomial}
B_{\textrm{chsh}}=ab+ab'+a'b-a'b',
\end{align}
 The product expectation values are easily obtained from the correlations, e.g., $\av{ab}= P(+1,+1|a,b) + P(-1,-1|a,b)-P(+1,-1|a,b) -P(-1,+1|a,b)$, etc.  
 
 \subsubsection*{Local correlations}
 \noindent\citet{chsh} showed that all local correlations obey the tight bound
 \begin{align}\label{chshineqintro}
|\av{B_{\textrm{chsh}}}_{\textrm{lhv}}|=|\av{ab}_{\textrm{lhv}} +\av{ab'}_{\textrm{lhv}}
 +\av{a'b}_{\textrm{lhv}}  -\av{a'b'}_{\textrm{lhv}}|
\leq2.
\end{align}
The local polytope is the subset in the four dimensional real space $\mathbb{R}^4$ of all vectors  $(\av{ab}, \av{ab'}, \av{a'b},  \av{a'b'})$\forget{(i.e., whose elements are all product expectation values)} that can be attained by local correlations. It is the convex hull in $\mathbb{R}^4$ of the 8 extreme points (vertices) that are of the form \begin{align}\label{extrloc}
(1,1,1,1),(-1,-1,-1,-1),(1,1,-1,-1),(-1,-1,1,1),\nn\\(1,-1,1,-1),(-1,1,-1,1),(1,-1,-1,1),(-1,1,1,-1).
\end{align}
 This polytope is the four-dimensional octahedron and has 8 trivial facets as well as 8 non-trivial ones. The trivial ones are the inequalities of the form 
 \begin{align}\label{triv}
 -1\leq\av{ab}_{\textrm{lhv}}\leq1 ,~~~~
  -1\leq\av{ab'}_{\textrm{lhv}}\leq 1,\nn\\
   -1\leq\av{a'b}_{\textrm{lhv}}\leq 1,~~~~
    -1\leq\av{a'b'}_{\textrm{lhv}}\leq 1.
  \end{align}The non-trivial facets are all equivalent to the CHSH inequality (\ref{chshineqintro}), up to trivial symmetries, giving a total of 8 equivalent inequalities, as first proven by \citet{fine}, cf. \citet{collinsgisin}. These eight are \cite{barrett05}: 
  \begin{align}\label{8chsh}
(-1)^\gamma \av{ab}_{\textrm{lhv}} +  (-1)^{\beta +\gamma}\av{ab'}_{\textrm{lhv}} +
(-1)^{\alpha +\gamma} \av{a'b}_{\textrm{lhv}} +(-1)^{\alpha+\beta+\gamma+1}\av{a'b' }_{\textrm{lhv}}\leq 2,
\end{align} with $\alpha,\beta,\gamma \in \{0,1\}$. These are the necessary and sufficient conditions for a LHV model to exist.

 \subsubsection*{Quantum correlations}\noindent
 In terms of the CHSH polynomial a non-trivial tight quantum bound is given by the Tsirelson inequality \cite{cirelson} 
\begin{align}\label{tsirelsonintro}
|\av{B_{\textrm{chsh}}}_{\textrm{qm}}| \leq 2\sqrt{2},
\end{align}
which can be reached by entangled states.
This shows that the local polytope is strictly contained in the quantum body, which can be regarded a concise statement of Bell's theorem \cite{bell64}.  Much more can be said about the structure of quantum violations of Bell-type inequalities, even in the simplest CHSH case, see e.g., \cite{tradeoff}, but we will refrain from doing that here. Our focus is on the no-signaling correlations, to which we turn next.

 \subsubsection*{No-signaling correlations}\noindent
No-signaling correlations are able to violate the Tsirelson inequality (\ref{tsirelsonintro}). A well known example of this is the joint distribution known as the Popescu-Rohrlich distribution \cite{prbox}, also known as the PR box that gives $\av{B_{\textrm{chsh}}}_{\textrm{ns}} = 4$, which is the absolute maximum $|B_{\textrm{chsh}}|_{\textrm{max}}$. In fact, it is an extreme point of the no-signaling polytope for the case of two dichotomous observables per party. Furthermore, all the no-signaling extreme points of this polytope have a such a form. They can all be written as \cite{barrett05}
\begin{align}\label{extremens}
P(A,B|a,b)  = \left \{ \begin{array}{l} 1/2 ,~~\textrm{if}~~A\oplus B=ab , \\  0,~~\textrm{otherwise}, \end{array} \right.
\end{align}
where $\oplus$ denotes addition modulo $2$. Here the outcomes $A,B$ and the settings $a,b$ are labeled by $0$ and $1$ respectively\forget{, i.e., $a_1,a_2,A_1,A_2\in\{0,1\}$}, where $0$ corresponds to outcome $+1$ and the unprimed observable respectively; and $1$ corresponds to outcome $-1$ and the primed observable respectively. 

There is a one-to-one correspondence between the non-local extreme points and the facets of the local polytope that are given by the CHSH inequalities  \eqref{8chsh}. To show this we note that the CHSH inequalities in the larger 16-dimensional space of correlations are equal to:
\begin{align}\label{chshlarger}
1\leq P(A=B)+P(A=B')+P(A'=B)+P(A'\neq B')\leq3
\end{align}
where  $P(A=B):= P(+1,+1|a,b)+P(-1,-1|a,b)$, and $P(A'\neq B'):= P(+1,-1|a',b')+P(-1,+1|a',b')$, etc.
  This gives two inequalities and the other six are obtained by permuting  the  primed and unprimed quantities for system 1 and 2 respectively. A total of 8 local extreme points saturate each of these inequalities. They are deterministic, i.e., $P(+1,+1|a,b)=P(+1|a)P(+1|b)$, etc., where $P(+1|b)$ and $P(+1|b)$ are either $0$ or $1$. 
Because these 8 extreme points are also linearly independent the inequalities \eqref{chshlarger} (and the equivalent ones) give the facets of the 8-dimensional local polytope in the larger space of correlations.

The 8 local extreme points that lie on each of the local facets are also extreme points of the no-signaling polytope. Only one extreme no-signaling correlation \eqref{extremens} is on top of each local facet, and it violates the CHSH inequality associated to this local facet maximally\forget{
its associated CHSH inequality is maximally violated by this no-signaling correlation} \cite{barrett05}. This is the one-to-one correspondence referred to above. 


The non-trivial facets of the no-signaling polytope are given by the defining equalities on the left hand side of \eqref{22nosig} and read in the dichotomic case 
\begin{align}\label{2nosig2}
\sum_{B=+1,-1}P(A,B|a,b)&=\sum_{B=+1,-1}P(A,B|a,b'),
\end{align}
for $a_1=+1,-1$, and analogous equalities are obtained by permutations of settings and outcomes so as to give a total of eight equalities.

In terms of expectation values we obtain non-trivial inequalities for the marginals\footnote{\label{wrongsignaling}In case 
 no-signaling obtains we can define $\av{a}_{\textrm{ns}}:=\av{a}_{\textrm{ns}}^{b}=\av{a}_{\textrm{ns}}^{b'}$  because the marginal for party 1 does not depent on the setting chosen by party 2 (cf. \eqref{22nosig}).  Inserting this in \eqref{Mnos} gives the trivial inequalities
$\av{a}_{\textrm{ns}}\leq\av{a}_{\textrm{ns}}$ and $\av{a}_{\textrm{ns}}\geq\av{a}_{\textrm{ns}}$. However, this misses the point. Because the non-trivial tight no-signaling Bell-type inequalities are supposed to discern the no-signaling correlations from more general correlations one must allow for the most general framework in which signaling is in principle possible., i.e, where the marginals can depend on the settings corresponding to the outcomes that are no longer considered.  This cannot be excluded from the start.}:
\begin{align}
\label{Mnos}
\av{a}_{\textrm{ns}}^{b}\leq\av{a}_{\textrm{ns}}^{b'} , ~~~\textrm{and} ~~~ \av{a}_{\textrm{ns}}^{b}\geq\av{a}_{\textrm{ns}}^{b'},
\end{align}
where we have used $\av{a}_{\textrm{ns}}^{b}:=\sum_{A} A~P(A|a)^{b}$  and $P(A|a)^{b}$ as defined in \eqref{M1} and obeying the  no-signaling constraint \eqref{22nosig}.

If we consider product expectation values instead of the marginal ones we only obtain trivial inequalities. In the space $\mathbb{R}^4$ of vectors with components $(\av{ab}, \av{ab'},\av{a'b}, \av{a'b'})$ the 8 no-signaling extreme points \eqref{extremens} give the following vertices   \begin{align}\label{extrns}
(-1,1,1,1),(1,-1,-1,-1),(1,-1,1,1),(-1,1,-1,-1),\nn\\(1,1,-1,1),(-1,-1,1,-1),(1,1,1,-1),(-1,-1,-1,1).
\end{align} In this space the no-signaling polytope is the convex hull of the 16 local extreme points (\ref{extrloc}) and of those given by (\ref{extrns}). Its facet inequalities are just the 8 trivial inequalities in (\ref{triv}) and therefore it is in fact just the four-dimensional unit cube \cite{pitowsky08}. 
We thus obtain only trivial facet inequalities.

In the next section we derive non-trivial no-signaling inequalities in terms of the product and marginal expectation values. Although these cannot be tight inequalities, i.e., they cannot be facets of the no-signaling polytope, we show them to do useful work nevertheless.
%
In order to obtain these inequalities we will have to consider a larger dimensional space than the four-dimensional of vectors $(\av{ab}, \av{ab'},\av{a'b}, \av{a'b'})$.

\subsubsection*{Comparing the different correlations}
\noindent
One of the no-signaling correlations (\ref{extremens}) was discovered already in 1985 independently by \citet{khalfin} and \citet{rastall} who also showed it to give the algebraic maximum for the CHSH expression. However, \citet{prbox} presented this correlation in order to ask an interesting question, not asked previously: Why do quantum correlations not violate the CHSH expression by a larger amount? Such a larger violation would be compatible with no-signaling, so why is quantum mechanics not more non-local? This paper by \citet{prbox} marked the start of a new research area, that of investigating no-signaling distributions and their relationship to quantum mechanics. Unfortunately we cannot review this exciting field here; see e.g. \cite{masanes06}. 

For the bi-partite case and two dichotomous observables per party the above results show how the different sets of correlations are related: Since some quantum correlations turn out to be non-local in the sense of not being of the local form, the set of quantum correlations is a proper superset of the set of local correlations. But it is a proper subset of the set of no-signaling correlations which are able to violate the Tsirelson inequality up to the absolute maximum.  
In summary, the CHSH polynomial gives inequalities that give a non-trivial tight bound for local and quantum correlations but not so for no-signaling correlations. Indeed,  the latter can reach the absolute maximum  $|B_{\textrm{chsh}}|_{\textrm{max}}$.

\section{IV: Discerning no-signaling correlations}\label{discerningno-signalingsection}
\noindent
In this section we search for non-trivial constraints on the expectation values that are a consequence of no-signaling. We derive a non-trivial  Bell-type inequality for the no-signaling correlations in terms of both product and marginal expectation values. It thus discerns such correlations from more general correlations. Although the inequalities do not indicate facets of the no-signaling polytope we show that they can provide interesting results nevertheless. They provide  constraints on no-signaling correlations that are required to reproduce the perfectly correlated and anti-correlated quantum predictions of the singlet state.

Before we present our new inequalities, we first take a look at a previous attempt to formulate such a set of non-trivial inequalities, and that we show to be flawed.

\subsection[The Roy-Singh no-signaling Bell-type inequality is trivially true]{The Roy-Singh no-signaling Bell-type inequality is\\ trivially true}
\noindent
\citet{roysingh} claimed to have obtained a non-trivial no-signaling Bell-type inequality in terms of expectation values. They assumed no-signaling  by requiring that the expectation value of the observable corresponding to setting $a$ only depends on this setting and not on the faraway setting $b$, and vice versa. Thus $\av{a}_\textrm{ns}=f(a)$ and  $\av{b}_\textrm{ns}=g(b)$ where $f$ and $g$ are some functions\footnote{This notation by Roy and Singh is awkward since it suggests that the expectation value solely depends on the setting and not also on the state of the system that is being measured. However, this is not the case since they in fact use the definition $\av{a}_{\textrm{ns}}:=\int d \lambda \rho(\lambda,a,b)A(\lambda,a,b)$, that incorporates the hidden-variable distribution of the system under consideration, and where the dependence on $b$ on the left hand side is left out because of no-signaling.}. The inequalities of Roy and Singh \cite{roysingh} read:
\begin{align}
|\,\av{ab}_\textrm{ns} \pm \av{a}_\textrm{ns}\,|&\leq 1\pm \av{b}_\textrm{ns}, \label{roy1}\\
|\,\av{ab}_\textrm{ns} \pm \av{b}_\textrm{ns}\,|&\leq 1\pm \av{a}_\textrm{ns} \label{roy2}.
\end{align}
Roy and Singh interpret their inequalities as testing theories that obey no-signaling against more general signaling theories, i.e., their inequalities are supposed to give a non-trivial bound for no-signaling correlations. It should be noted that \citet{avis} also  claim that this set of inequalities gives a nontrivial no-signaling bound (see their Proposition 2 in \cite{avis}). But falsely so, we claim.

We mention two points of criticism; the first minor, the second major: First, one should  include the far-away setting in the marginals expectation values (i.e., use $\av{a}_\textrm{ns}^{b}$ and $\av{b}_\textrm{ns}^{a}$), as was argued in footnote \ref{wrongsignaling}.  Secondly, and more importantly, no correlation whatsoever can violate these inequalities, whether they are signaling or not.  The inequalities are trivially true and are therefore irrelevant. The reason for this is that they follow from the trivial constraint that the probabilities $P(a,b|A,B)$ are non-negative. Let us show why this is the case.

The Roy-Singh inequalities \eqref{roy1} and \eqref{roy2} are in fact equivalent to the set of inequalities 
\begin{align}\label{seteq}
-1+|\av{a}^b+\av{b}^a|\leq \av{ab}\leq 1-|\av{a}^b-\av{b}^a|
\end{align}
that can be easily shown to hold for any possible correlation. Note that we leave out the subscript `ns', but include in the marginal expectation values $\av{a}^b$, $\av{b}^a$ the setting at the other side because there might be a dependency on the far-away setting as we are no longer restricting ourselves to no-signaling correlations.

 The inequality \eqref{seteq} was first derived by \citet{leggett} in the following way (cf. \cite{paterek,branciard}).
For quantities $a,b$ that can take outcomes $A=\pm1$ and $B=\pm1$  the following identity holds:
\begin{align}\label{outcomesid}
-1+|A+B|=AB=1-|A-B| .
\end{align}
Let the outcome $A$ be determined\footnote{Without any further constraints, it is mathematically always possible to let the outcomes be determined by a deterministic hidden variable model.} by some hidden variable $\lambda$ and by the settings $a,b$: \mbox{$A:=A(\lambda,a,b)$}.  Furthermore, let  $\av{a}^b:= \int_\Lambda d\lambda \mu(\lambda|a,b) A(\lambda,a,b)$ be the average of quantity $A$ with respect to some positive normalized weight function $ \mu(\lambda|A,B)$ over the hidden variables. This function can contain any non-local or signaling dependencies on the setting $a$ and $b$. Define similarly the quantity $\av{b}^a$ and the average of the product $AB$ denoted by $\av{ab}$. Taking the average of the expression in \eqref{outcomesid} and using the fact that the average of the modulus is greater or equal to the modulus of the averages  one obtains the set of inequalities \eqref{seteq}.

Although the Roy-Singh inequalities indicate that the marginals  $\av{a}^b$ and  $\av{b}^a$ are not independent of the product expectation value $\av{ab}$, and vice versa, this is only a consequence of non-negativity of joint probabilities and not of the requirement of no-signaling. 
In conclusion, the Roy-Singh inequalities (and consequenty also those by \citet{avis}) fail to show what they were supposed to do.
 However, we next present a derivation that does meet this task of providing a non-trivial no-signaling Bell-type inequality in terms of both product and marginal expectation values.

\subsection{Non-trivial  
 no-signaling Bell-type inequalities}\label{nontrivnosignal}
 \noindent
Recall that the CHSH inequality does not suffice for discerning no-signaling correlations from general correlations because no-signaling correlations can reach the absolute maximum of this inequality. Indeed, using only product expectation values it was shown that the no-signaling polytope in the corresponding 4-dimensional space of vectors with components $\av{ab}, \av{ab'},\av{a'b}, \av{a'b'}$ is the trivial unit-cube. Our analysis must thus be performed in a larger space, and we consider the vectors that have as components  in addition to the product expectation values the marginal ones, i.e., we also consider the quantities $\av{a}^b, \av{a}^{b'},\av{a'}^b$, etc. In this space we obtain a set of non-trivial no-signaling Bell-type inequalities that discerns the no-signaling correlations from more general correlations.

The trick we use to obtain the new set of inequalities is to combine two different Roy-Singh inequalities (see \eqref{roy1} and \eqref{roy2}) where the no-signaling constraint is invoked to
set  $\av{a}_\textrm{ns}^b=\av{a}_\textrm{ns}^{b'}:=\av{a}_{\textrm{ns}}$, etc. 

For example, consider the following two Roy-Singh inequalities that hold for all correlations: 
\begin{align}
|\,\av{ab} \pm \av{a}^b\,|&\leq 1\pm \av{b}^a, \\
|\,\av{a'b} \pm \av{a'}^b\,|&\leq 1\pm \av{b}^{a'}.
\end{align}
Using  the inequality $|x+y|\leq |x|+|y|$ ($x,y\in\mathbb{R}$) we obtain
\begin{align}
|\av{ab} +\av{a}^b +\av{a'b} -\av{a'}^b|\leq |\av{ab} +\av{a}^b| +|\av{a'b} -\av{a'}^b|
\leq 2+\av{b}^a-\av{b}^{a'}.
\end{align}
Assuming no-signaling (i.e., we set  $\av{b}_\textrm{ns}^a=\av{b}_\textrm{ns}^{a'}:=\av{b}_{\textrm{ns}}$) gives\footnote{That this is non-trivial can be shown by  giving an example of a signaling correlation that violates \eqref{nosignontriveq}. Consider a deterministic protocol where if $a$ and $b$ are measured jointly party 1 obtains outcome  $A_{11}$ and party 2 obtains outcome $B_{11}$, and, alternatively, if $a'$ and $b$ are measured jointly party 1 obtains outcome  $A_{21}$ and party 2 obtains outcome $B_{21}$, where $B_{11}\neq B_{21}$.
Then $\av{ab}=A_{11}B_{11}$, $\av{a'b}=A_{21}B_{21}$, $\av{a}^b=A_{11}$,  $\av{a'}^b=A_{21},\av{b}^a=B_{11},\av{b}^{a'}=B_{21}$. This is a one-way signaling protocol because $\av{b}^a\neq \av{b}^{a'}$.  If one chooses $A_{11}=B_{11}=1$ and $A_{21}=B_{21}=-1$ a value of 4 is obtained for the left hand-side of \eqref{nosignontriveq} clearly violating this inequality.}:
\begin{align}\label{nosignontriveq}
|\av{ab}_{\textrm{ns}} +\av{a'b}_{\textrm{ns}} +\av{a}_{\textrm{ns}}^b -\av{a'}_{\textrm{ns}}^b|\leq 2.
\end{align}

A total of 32 different such inequalities can be obtained that we can write as 
\begin{subequations}\label{nontrivnonsig2}
\begin{align}
(-1)^\gamma \av{ab}_{\textrm{ns}} +(-1)^{\beta +\gamma}\av{a'b}_{\textrm{ns}} +(-1)^{\alpha +\gamma} \av{a}_{\textrm{ns}}^B +(-1)^{\alpha+\beta+\gamma+1}\av{a'}_{\textrm{ns}}^B\leq 2,\\
(-1)^\gamma \av{ab}_{\textrm{ns}} +(-1)^{\beta +\gamma}\av{ab'}_{\textrm{ns}} +(-1)^{\alpha +\gamma} \av{b}_{\textrm{ns}}^A +(-1)^{\alpha+\beta+\gamma+1}\av{b'}_{\textrm{ns}}^A\leq 2,\\
(-1)^\gamma \av{a'b'}_{\textrm{ns}} +(-1)^{\beta +\gamma}\av{a'b}_{\textrm{ns}} +(-1)^{\alpha +\gamma} \av{b}_{\textrm{ns}}^{a'} +(-1)^{\alpha+\beta+\gamma+1}\av{b'}_{\textrm{ns}}^{a'}\leq 2,\\
(-1)^\gamma \av{a'b'}_{\textrm{ns}} +(-1)^{\beta +\gamma}\av{ab'}_{\textrm{ns}} +(-1)^{\alpha +\gamma} \av{a}_{\textrm{ns}}^{b'} +(-1)^{\alpha+\beta+\gamma+1}\av{a'}_{\textrm{ns}}^{b'}\leq 2,
\end{align}
\end{subequations}
with $\alpha,\beta,\gamma \in \{0,1\}$.

If we compare these inequalities to the CHSH inequality  $|\av{ab}_{\textrm{lhv}} +\av{a'b}_{\textrm{lhv}} +\av{ab'}_{\textrm{lhv}} -\av{a'b'}_{\textrm{lhv}}|\leq 2$ for local correlations we see a remarkable structural similarity: we only have to replace two product expectation values by some  specific marginal expectation values.

Adding two different Roy-Singh inequalities  and assuming no-signaling gives a slightly different  inequality that contains six terms\footnote{This is indeed non-trivial. The deterministic signaling protocol where the outcomes are $A_{11}=A_{22}=-1$ and $A_{12}=B_{12}=A_{21}=B_{21}=B_{11}=B_{22}=1$ 
gives $\av{ab}=A_{11}B_{11}=-1$, $\av{a'b'}=A_{22}B_{22}=-1$, and  $\av{a}^{b'}=A_{12}=1,\av{a'}^{b}=A_{21}=1,\av{b}^{a'}=B_{21}=1,\av{b'}^a=B_{12}=1$ so as to give a value of 6 on the left hand side of \eqref{nsres333} and which violates this inequality.}:
 \begin{align}\label{nsres333}
 -\av{ab}_\textrm{ns}-\av{a'b'}_\textrm{ns}+\av{a}_\textrm{ns}^{b'}+\av{b}_\textrm{ns}^{a'}+\av{a'}_\textrm{ns}^{b}+\av{b'}_\textrm{ns}^{a}\leq 2
\end{align}
Using permutations of observables and outcomes in \eqref{nsres333}\footnote{There are 6 different permutations that are of two types:  3 different permutations of the outcomes:  for party 1, for party 2 and for both parties; and 3 different permutations for the observables: permute $a$ with $a'$, $b$ with $b'$ or perform both permutations at once. All different combinations of these six give 64 possibilities of which only 14 give distinct non-trivial inequalities.}
  a total of 14 different non-trivial inequalities  can be obtained. These  can be compactly written as
\begin{subequations}\label{14ineq}
\begin{align}\label{14ineqa}
&-\av{ab}_\textrm{ns}-\av{a'b'}_\textrm{ns}-(-1)^\alpha\av{a}_\textrm{ns}^{b'}-(-1)^\alpha\av{b}_\textrm{ns}^{a'}-
\nn\\&~~~~~~~~~~~~~~~~~~~~~~~~~~~~~~~~~~~~~~~~~
(-1)^\beta\av{a'}_\textrm{ns}^{b}-(-1)^\beta\av{b'}_\textrm{ns}^A\leq 2,\\
&-(-1)^\gamma\av{ab}_\textrm{ns}-(-1)^{\gamma+1}\av{a'b'}_\textrm{ns}-(-1)^{1+\gamma\delta}\av{a}_\textrm{ns}^{b'}-(-1)^{1-\gamma(\delta+1)}\av{b}_\textrm{ns}^{a'}\nn\\&~~~~~~~~~~~~~~~~~~~~~~~-(-1)^{(\delta+1)(1-\gamma)+1}\av{a'}_\textrm{ns}^B-(-1)^{1+\delta(1-\gamma)}\av{b'}_\textrm{ns}^A\leq 2,\label{14ineqb}
\end{align}
\end{subequations}
where $\alpha,\beta,\gamma,\delta\in \{0,1\}$ except for the case $\alpha=\beta=0$ which is excluded since it gives a trivial inequality (see \eqref{result2}). This specifies 7 inequalities and the other 7 are obtained by interchanging $A$ by $A'$.

None of the above no-signaling inequalities are facetsof the no-signaling polytope. They are saturated by only 7 affinely independent extreme points instead of the required 8 which is necessary for a facet.


\subsubsection{No-signaling perfect correlations for identical systems implies local randomness} 
\noindent
The set of non-trivial inequalities \eqref{14ineq} can be used to prove the following result: if different parties use identical devices 
and have perfect correlations at spacelike separation between dichotomic observables, then because of no-signaling the local marginals cannot but be completely random. In other words, the existence of no-signaling perfect correlations for identical systems implies local randomness. This is proven as follows.

\forget{
The set of non-trivial inequalities \eqref{14ineq} shows an interesting constraint on no-signaling correlations that are required to reproduce the perfectly correlated and anti-correlated quantum predictions of the two-qubit singlet state $(\ket{01}-\ket{10})/\sqrt{2}$.
}

Consider spin measurements in directions $\vec{a}$ and $\vec{b}$ performed by parties $I$ and $II$ respectively. Suppose that they are able to establish perfect anti-correlated predictions when the measurements are in the same direction, and perfect correlated predictions when they are in opposite directions\footnote{Such correlations are, for example, present in the two-qubit singlet state $(\ket{01}-\ket{10})/\sqrt{2}$.}:  
\begin{align}
\forall \,\vec{a},\,\vec{b}:~~\av{\vec{a}\vec{b}}=-1,~~\textrm{when} ~~\vec{a}=\vec{b}, \label{perfectanticorre}\\
\forall \,\vec{a},\,\vec{b}:~~\av{\vec{a}\vec{b}}=1,~~\textrm{when} ~~\vec{a}=-\vec{b}.\label{perfectcorre}
\end{align} 
Suppose one wants to reproduce these correlations using a no-signaling correlation, i.e., for all $\vec{a},\vec{a}',\vec{b},\vec{b}'$ inequalities \eqref{14ineqa} and \eqref{14ineqb} for all admissible $\alpha,\beta,\gamma,\delta$ must hold, where the settings $a,a',b,b'$ have been denoted by  
the vectors  $\vec{a},\vec{a}',\vec{b},\vec{b}'$ respectively.   Because of no-signaling the dependence of the marginals on far-away settings is dropped,  i.e., $\av{\vec{a}}^{\vec{b}}=\av{\vec{a}}^{\vec{b}'}:=\av{\vec{a}}$, etc.

In the case where $\vec{a}'=\vec{b}=\vec{b}'=\vec{a}$ the assumption \eqref{perfectanticorre}  together with the constraint \eqref{14ineqa} for $\alpha=\beta=1$ implies, for all $\vec{a}$:
\begin{align}\label{result1}
-\av{\vec{a}}^I_\textrm{ns}-\av{\vec{a}}^{II}_\textrm{ns}\geq 0.
\end{align}
where the two different parties $I$ and $II$ are explicitly indicated, i.e., $\av{\vec{a}}^I_\textrm{ns}$ for party $I$ and $\av{\vec{a}}^{II}_\textrm{ns}$ for party $II$.

 Furthermore, non-negativity gives  $4P(++|\vec{a}\vec{b})+4P(++|\vec{a}'\vec{b}')\geq0$, which is identical to
  \begin{align}\label{result2}
 \av{\vec{a}\vec{b}}_\textrm{ns}+ \av{\vec{a}'\vec{b}'}_\textrm{ns}+\av{\vec{a}}_\textrm{ns}+\av{\vec{b}}_\textrm{ns}+\av{\vec{a}'}_\textrm{ns}+\av{\vec{b}'}_\textrm{ns}+2 \geq0.
 \end{align}  
 In the case where $\vec{a}'=\vec{b}=\vec{b}'=\vec{a}$  assumption \eqref{perfectanticorre}  and the constraint \eqref{result2} imply, for all 
 $\vec{a}$: $\av{\vec{a}}^I_\textrm{ns}+\av{\vec{a}}^{II}_\textrm{ns}\geq 0$.
 Together with \eqref{result1} we thus obtain, for all $\vec{a}$:
\begin{align}\label{result4}
\av{\vec{a}}^I_\textrm{ns}+\av{\vec{a}}^{II}_\textrm{ns}= 0.
\end{align}
This is the first non-trivial constraint.

The second constraint follows from the case where $-\vec{a}'=\vec{b}=\vec{b}'=\vec{a}$. In this case the assumption \eqref{perfectcorre}  together with the constraints of \eqref{14ineqb} for $\gamma=\delta=0$ and $\gamma=0,\delta=1$ implies, for all $\vec{a}$: $\av{\vec{a}}^I_\textrm{ns}=\av{-\vec{a}}^{II}_\textrm{ns}$. Together with \eqref{result4} this implies, for all $\vec{a}$:
\begin{align}\label{result5}
\av{-\vec{a}}^I_\textrm{ns}=-\av{\vec{a}}^I_\textrm{ns}.
\end{align}
By symmetry the same holds for party $II$. 

Thus \eqref{result4} and \eqref{result5} are necessary conditions for any no-signaling model to reproduce the perfect (anti-)correlations of \eqref{perfectanticorre}, \eqref{perfectcorre}. These conditions state that the marginal expectation values for party $I$ and $II$ must add up to zero  for measurements in the same direction, and the individual marginal expectation values must be odd functions of the settings.
Consequently, any model reproducing the perfect (anti-) correlations and  which does not obey either one (or both) of these conditions must be signaling.

In case the no-signaling model treats the systems held by party $I$ and $II$ the same, i.e., $\av{\vec{a}}^I_\textrm{ns}=\av{\vec{a}}^{II}_\textrm{ns}$ (or, equivalently: $P(A=+|\vec{x})=P(B=+|\vec{y})$ for all $\vec{x}=\vec{y}$), it must have vanishing marginal expectation values: $\av{\vec{a}}^I_\textrm{ns}=\av{\vec{a}}^{II}_\textrm{ns}=0$.  All marginal probabilities then must be uniformly distributed: for all $\vec{a}$, $P(+|\vec{a})=P(-|\vec{a})=\frac{1}{2}$, etc.\footnote{In case one requires not only the perfect (anti-) correlations for parallel and anti-parallel settings but the full singlet state correlation $\av{\vec{a}\vec{b}}=-\vec{a}\cdot\vec{b}$, $\forall\,\vec{a},\vec{b}$, the requirement of vanishing marginal expectation values must indeed obtain. \citet{branciard} established this for hidden-variable models of the Leggett type, but it holds also for general no-signaling models.} 

Note that if one makes the natural assumption that measuring $\vec{a}$ is equivalent to measuring $-\vec{a}$ but with opposite outcomes, i.e., $P(-|\vec{a})=P(+|-\vec{a})$, then one does not need to assume $P(A\neq B|\vec{a}=\vec{b})=1$ (i.e., perfect anticorrelation when measured in the same directions, see \eqref{perfectcorre}). Requiring $P(A= B|\vec{a}=-\vec{b})=1$ (i.e., perfect correlation when measured in opposite directions,  see \eqref{perfectanticorre}) then suffices\footnote{This was remarked by Nicolas Gisin, priviate communication.}.

%
%

\section{V: Discussion: nonlocal no-signaling correlations imply secrecy}

It can be easily seen that uniformly random local outputs guarentees no-signaling (there can be no dependence an anything, let alone one the far-away setting). The opposite does not hold in general as no-signaling correlations exist that have biased local marginals.  However,
we have shown here that if the two different parties use identical devices and have perfect correlations at spacelike separation between dichotomic observables, then because of no-signaling the local marginalsmust be completely random. In other words, the existence of no-signaling perfect correlations for identical systems implies local randomness. It would be interesting to design a
protocol that implements this type of local randomness via
non-local correlations and that has cryptographic purposes
such as secure key distribution\footnote{How can it be possible to derive secrecy from correlations alone? 
Crucial is that correlations can be monogamous to some extent~\cite{seevinckmonogamy}. For example, if Alice and Bob share a maximally entangled bi-partite quantum state, then Eve cannot be correlated to either Alice's or Bob's subsystem. Classically, however, no such an effect is known: If Alice and Bob have highly correlated bits, Eve can nevertheless obtain them.   

Such monogamous correlations are always non-local in the sense that they violate a Bell-type inequality. For quantum correlations this implies that the idea that quantum physics is incomplete and should be augmented by classical variables determining the behavior of every system under any possible measurement just does not work: these variables do not exist. This holds true for all non-local correlations, including no-signaling ones. This is what can be exploited cryptographically: if such variables do not exist, then no adversary could have known them beforehand so as to eavesdrop.
}.
\forget{The outputs are, hence, perfectly random and the randomness must have been generated after
input reception. 
This is what we can make use of for key agreement: Assume that Alice and Bob share any kind of
 physical system, carry out space-like separated measurements (hereby excluding message transmission), 
and measure data having the perfect correlations as described above. (In order to test this, they exchange all 
the input bits and some randomly chosen outputs.) The resulting data are then perfectly secret bits, 
because even conditioned on an adversary's complete information, the correlation between Alice and 
Bob must be non-signaling.
}


\begin{theacknowledgments}The author thanks Nicolas Gisin for his encouraging remarks on presenting the novel result at the end of section IV.
\end{theacknowledgments}

\bibliographystyle{aipproc}   

\end{document}